\documentclass{article}

\usepackage{arxiv}

\usepackage[utf8]{inputenc} 
\usepackage[T1]{fontenc}    

\usepackage{url}            
\usepackage{booktabs}       
\usepackage{nicefrac}       
\usepackage{microtype}      
\usepackage{lipsum}
\usepackage{graphicx}
\usepackage{amsmath,amsfonts,amssymb}
\usepackage[colorlinks=true, allcolors=blue]{hyperref}
\usepackage{float}
\usepackage{subcaption}
\usepackage{changepage}

\graphicspath{ {./images/} }

\title{Application of the latent twins approach for clear sky retrieval from IASI observations}

\author{
 Michele Martinazzo \\
  Department of Physics and Astronomy Augusto Righi\\
  University of Bologna\\
  Bologna, Italy \\
  \texttt{michele.martinazzo2@unibo.it} \\
   \And
 Cristina Sgattoni \\
  Institute of BioEconomy (IBE) Firenze\\
  National Research Council \\
  Firenze, Italy \\
  \texttt{cristina.sgattoni@cnr.it} \\
  \And
 Marco Menarini \\
  Institute of Atmospheric Sciences and Climate\\
  National Research Council\\
  Bologna, Italy \\
  \texttt{m.menarini@isac.cnr.it} \\
  \And
 Chiara Zugarini \\
  Institute for Applied Mathematics\\
  National Research Council\\
  Sesto Fiorentino-Firenze, Italy. \\
  \texttt{chiara.zugarini@fi.iac.cnr.it} \\
  \And
 Tiziano Maestri \\
  Department of Physics and Astronomy Augusto Righi\\
  University of Bologna\\
  Bologna, Italy \\
  \texttt{tiziano.maestri@unibo.it} \\
  \And
 Luca Sgheri \\
  Institute for Applied Mathematics\\
  National Research Council\\
  Sesto Fiorentino-Firenze, Italy. \\
  \texttt{luca@fi.iac.cnr.it} \\
}

\begin{document}
\maketitle
\begin{abstract}
In recent years, data-driven approaches emerged as alternatives to traditional physics-based retrievals, taking advantage of machine learning techniques such as learnable pseudoinverse, random forests, or deep learning architectures. 
Classical data-driven models generalize poorly to out-of-sample regimes, as they optimize over finite datasets without incorporating underlying physical laws. This often requires large models and extensive data to achieve reliability. Physics-Informed Neural Networks address this by embedding physical constraints into the learning process, enabling improved extrapolation. However, they requires substantial computational cost due to the need to solve governing equations at each training step. 
In this work, we introduce a novel deep learning architecture, based on latent twin approach, that balances model complexity, dataset size, and training cost, while providing a quantitative measure of data quality.
This architecture is applied to IASI spectra, with the goal to assess the robustness of this method for retrieving atmospheric profiles, including temperature, water vapor, ozone, surface emissivity, and surface temperature, in real-world clear-sky conditions. The algorithm is first applied on synthetic radiances derived from the NWP SAF database using the fast radiative transfer code $\sigma$-IASI/F2N. 
After validating the architecture on synthetic data, the algorithm is applied to IASI Level 1C observations, along with their corresponding Level 2 products which serve as reference to evaluate the reconstruction accuracy of the autoencoder-based retrieval. The retrieval performances are discussed along with possible strategies to provide an error analysis for the reconstructed thermodynamical profiles.
\end{abstract}


\section{Introduction}
\label{sec:intro}  

Satellite observations of the atmosphere are now central to Earth system science. Their global, continuous coverage has greatly improved our ability to monitor, understand, and predict weather and climate.\\
In order to extract information from the radiation acquisitions, classical Optimal Estimation (OE) retrievals are widely employed. These methodologies ensure radiometric consistency between the retrieved parameters and observed radiances through explicit forward modeling, while the Bayesian framework allows for a straightforward characterization of the retrieval errors\cite{Rodgers1976,SERIO2024,MARTINAZZO2026}. On the other hand, the iterative nature of these algorithms makes them computationally expensive, with inversion times of the order of $10^2$ to $10^4$ seconds, depending on the complexity of the chosen forward model. Moreover, the standard OE formulation assumes Gaussian statistics for the measurement errors and the prior, so its uncertainty characterization is exact only for a linear forward model and otherwise relies on linearization around the solution.
In recent years, data-driven approaches have emerged as powerful complements to traditional physics-based retrievals. In this work, we present a clear-sky retrieval framework based on the latent twins artificial-intelligence architecture to derive atmospheric thermodynamic and composition profiles from satellite infrared-sounder radiances. The framework is applied to Infrared Atmospheric Sounding Interferometer (IASI)\cite{SIMEONI1997} observations to retrieve vertical profiles of temperature, water vapor, and ozone, together with surface temperature and emissivity. By learning the nonlinear mapping between spectra and atmospheric state from a representative training database, the latent twins framework enables nearly instantaneous, profile-level inference without the need for iterative radiative-transfer calculations at run time.
Latent twins architectures were initially introduced in theoretical works by Chung et al.\cite{Chung_2026,chung2026_conform}, where their mathematical properties were explored. Their application to radiative-transfer retrieval problems under all-sky conditions was later demonstrated by Sgattoni et al. \cite{Sgattoni_2026}, using synthetic radiances from the Far-infrared Outgoing Radiation Understanding and Monitoring (FORUM) mission\cite{palchetti20}. This previous work showed the model's capability for scene recognition and its viability as a basis for retrieving atmospheric properties under all-sky conditions.
Here, we adopt this framework, proposing good practices and a methodology for its application to clear-sky conditions, and assess its robustness on real observational data from IASI.
To the best of our knowledge, this is the first application of this class of architectures to real satellite data.

\section{Theoretical background and architecture}

\subsection{Architecture description}
\label{sec:Architecture_description}

\begin{figure} [ht]
\begin{center}
\begin{tabular}{c} 
\includegraphics[height=6.5cm]{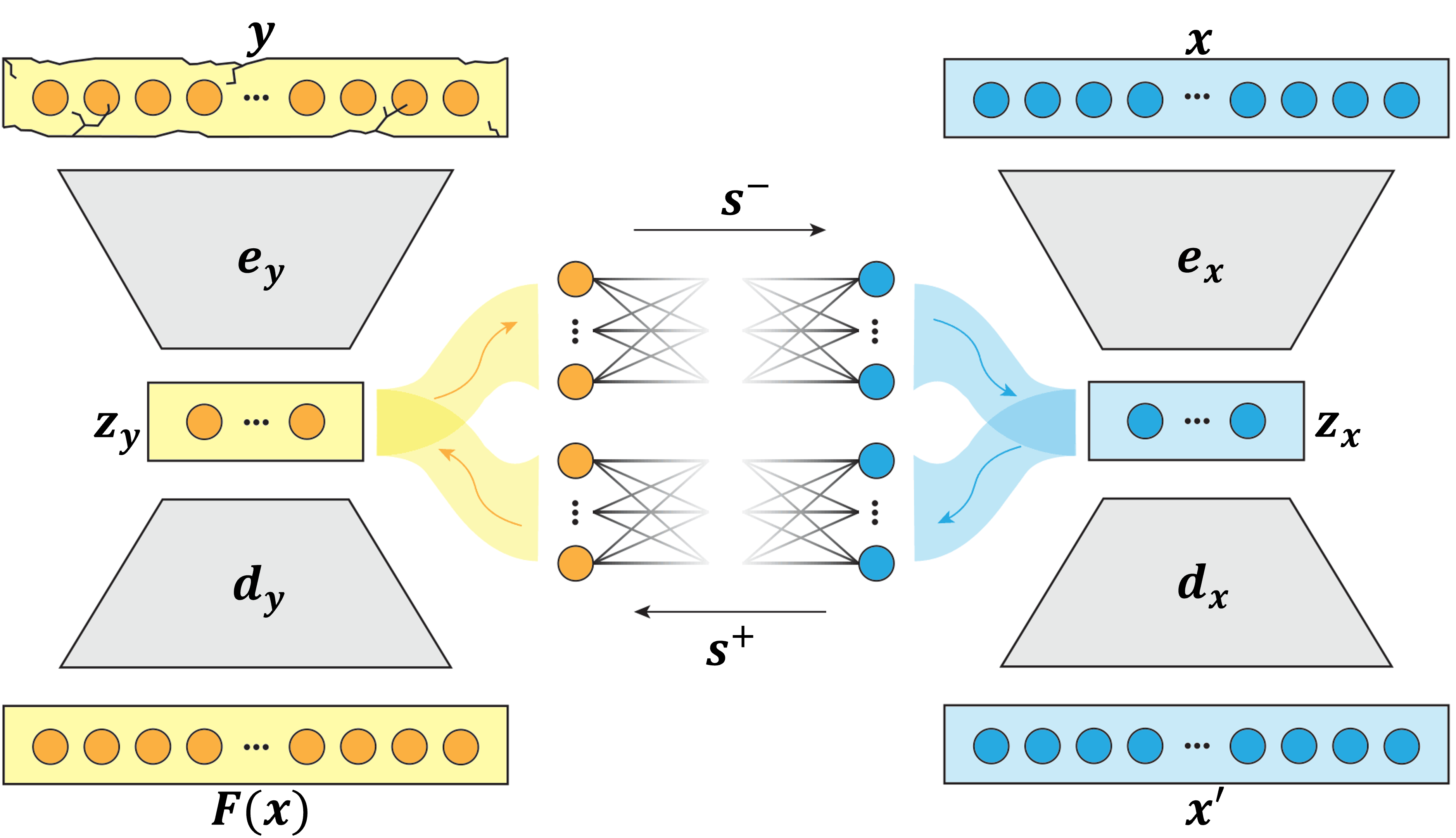}
\end{tabular}
\end{center}
\caption[LTarchitecture] 
   { \label{fig:LT_architecture} 
Schematic of the Latent Twin architecture for satellite inverse problems. Two autoencoders ($d_y\circ e_y$ and $d_x\circ e_x$) are jointly trained on the parameter space $x$ and the observation space $y$, producing low-dimensional latent representations $z_x$ and $z_y$, respectively. Simultaneously, forward and inverse operators are learned in latent space via trainable mappings $s^{+} : z_x \mapsto z_y$ and $s^{-} : z_y \mapsto z_x$. Additionally, the observation autoencoder is trained as a Denoising AutoEncoder (DAE), decoding the measurement latent representation into the theoretical noiseless signal $F(x)$. Both encoders and decoders comprise two hidden layers, whose widths decrease linearly towards the latent dimension; the decoders mirror this structure in reverse. The latent mappings consist of a single hidden layer of width equal to $1.2\,\dim(\mathcal{Z})$. ReLU activation functions are used.}
\end{figure}

The latent twins framework, as described by Chung et al.\cite{Chung_2026}, is a machine learning architecture designed to exploit the intrinsic low-dimensional structure present in high-dimensional data to solve inverse problems. It employs a pair of coupled autoencoders that learn compact representations of both the observed data and the corresponding quantities of interest, mapping each into its own lower-dimensional latent space. By learning the mappings between these two latent spaces, the framework constructs a surrogate model capable of approximating both the forward and inverse processes.
Compared to traditional data-driven approaches, this framework achieves significant performance gains, delivering robust estimates of the recovered quantities of interest, as shown by Chung et al.\cite{Chung_2026} and Sgattoni et al. \cite{Sgattoni_2026}. Moreover, although not used in this work, the possibility of using Variational AutoEncoders\cite{Kingma2019_Variational_AE} (VAE) enables uncertainty quantification for the inverse solution.\\
From an Earth observation point of view, we can link a satellite measurement $y\in \mathbb{R}^{m}$ with the observed quantity state $x\in \mathbb{R}^{n}$, through its forward problem:
\begin{equation}
y = F(x)+\epsilon
\end{equation}
where $F$ accounts for both the physical process and the measurement system, and $\epsilon$ denotes an additive measurement noise. In this framework, the latent twins is applied to solve the inverse problem, i.e. given a measurement $y$, reconstruct the unknown model parameters $x$.\\
A graphical description of the latent twins is provided in Fig.~\ref{fig:LT_architecture}.

The architecture consists of two autoencoders, one operating on the physical parameters and one on the observations. The first, $a_{\mathbf{x}} = d_{\mathbf{x}} \circ e_{\mathbf{x}}$, is defined by an encoder $e_{\mathbf{x}}: \mathbb{R}^{n} \rightarrow \mathcal{Z}_{\mathbf{x}}$ and a decoder $d_{\mathbf{x}}: \mathcal{Z}_{\mathbf{x}} \rightarrow \mathbb{R}^{n}$, where $\mathcal{Z}_{\mathbf{x}}$ denotes the latent space associated with the physical parameters. 
The second autoencoder, $a_{\mathbf{y}} = d_{\mathbf{y}} \circ e_{\mathbf{y}}$, is defined analogously by an encoder $e_{\mathbf{y}}: \mathbb{R}^{m} \rightarrow \mathcal{Z}_{\mathbf{y}}$ and a decoder $d_{\mathbf{y}}: \mathcal{Z}_{\mathbf{y}} \rightarrow \mathbb{R}^{m}$, with $\mathcal{Z}_{\mathbf{y}}$ the corresponding latent space for the observations.\\
The latent spaces $\mathcal{Z}_{\mathbf{x}}$ and $\mathcal{Z}_{\mathbf{y}}$ are 
chosen to have lower dimensionality than their respective input spaces, i.e., $\dim(\mathcal{Z}_{\mathbf{x}}) < n$ and $\dim(\mathcal{Z}_{\mathbf{y}}) < m$,  so that each autoencoder is forced to retain only the most informative structure 
in the data while discarding noise and redundancies. In the context of inverse problems, this bottleneck acts as an implicit regularization mechanism, promoting generalization and robustness to observational noise.\\
The two latent spaces are then coupled via a pair of trainable mappings, $s^{+}: \mathcal{Z}_{\mathbf{x}} \rightarrow \mathcal{Z}_{\mathbf{y}}$ and $s^{-}: \mathcal{Z}_{\mathbf{y}} \rightarrow \mathcal{Z}_{\mathbf{x}}$, which allow the network to learn both directions of the state-observation relationship. Of particular interest is the inverse surrogate

\begin{equation}\label{eq:inv_map}
    f^{-}(\mathbf{y}) \approx (d_{\mathbf{x}} \circ s^{-} \circ e_{\mathbf{y}})(\mathbf{y}),
\end{equation}
which approximates the solution to the inverse problem by mapping observed data back to the physical parameter space through the learned latent correspondence.\\
An important distinction with respect to \cite{Sgattoni_2026} is the adoption of a Denoising Autoencoder (DAE) for the observation branch of the model \cite{bengio2013}. Rather than learning a direct reconstruction $\mathbf{y} \rightarrow \mathcal{Z}_{\mathbf{y}} \rightarrow \mathbf{y}$, the encoder-decoder pair $(e_{\mathbf{y}}, d_{\mathbf{y}})$ is trained to recover clean observations from deliberately corrupted inputs, by minimizing
\begin{equation}
    \mathcal{L}_{\text{DAE}} = \mathbb{E}_{\mathbf{y}_{\epsilon}, \mathbf{y}} 
    \left[ \left\| d_{\mathbf{y}}(e_{\mathbf{y}}(\mathbf{y}_{\epsilon})) - 
    \mathbf{y} \right\|^2 \right],
\end{equation}
where $\mathbf{y}_{\epsilon}$ denotes a stochastically corrupted version of $\mathbf{y}$, and the expectation values is taken over paired samples. This encourages $\mathcal{Z}_{\mathbf{y}}$ to capture the intrinsic structure of the observations while remaining invariant to noise, which is particularly relevant in satellite-based Earth Observation where measurements are affected by multiple noise sources.


\subsection{Loss function and regularization}

The full model is trained end-to-end by minimizing a composite loss that combines the reconstruction objectives of both autoencoders with the consistency of the latent coupling. The total loss is defined as
\begin{equation}
    \mathcal{L} = \omega_{\mathbf{x}\mathbf{x}}\,\mathcal{L}_{\mathbf{x}\mathbf{x}} 
                + \omega_{\mathbf{y}_{\epsilon}\mathbf{y}}\,\mathcal{L}_{\mathbf{y}_{\epsilon}\mathbf{y}} 
                + \omega_{\mathbf{x}\mathbf{y}}\,\mathcal{L}_{\mathbf{x}\mathbf{y}} 
                + \omega_{\mathbf{y}_{\epsilon}\mathbf{x}}\,\mathcal{L}_{\mathbf{y}_{\epsilon}\mathbf{x}},
                + \omega_{\mathcal{R}}\,\mathcal{R},
\end{equation}
where the scalar weights $\omega_{\mathbf{x}\mathbf{x}}$, $\omega_{\mathbf{y}_{\epsilon}\mathbf{y}}$,$\omega_{\mathbf{x}\mathbf{y}}$, $\omega_{\mathbf{y}_{\epsilon}\mathbf{x}}$ and $\omega_{\mathcal{R}}$ control the relative contribution of each term. The first two terms are the autoencoder reconstruction losses,
\begin{equation}
    \mathcal{L}_{\mathbf{x}\mathbf{x}} = \mathbb{E}_{\mathbf{x}} 
    \left[ \left\| d_{\mathbf{x}}\big(e_{\mathbf{x}}(\mathbf{x})\big) - \mathbf{x} \right\|^2 \right],
    \qquad
    \mathcal{L}_{\mathbf{y}_{\epsilon}\mathbf{y}} = \mathcal{L}_{\text{DAE}},
\end{equation}
for the physical parameters and the observations, respectively, and where the expectation is taken over paired samples. The remaining two terms penalize inconsistencies in the latent coupling: $\mathcal{L}_{\mathbf{x}\mathbf{y}}$ measures the error of the forward mapping $s^{+}: \mathcal{Z}_{\mathbf{x}} \to \mathcal{Z}_{\mathbf{y}}$,
\begin{equation}
    \mathcal{L}_{\mathbf{x}\mathbf{y}} = \mathbb{E}_{\mathbf{x},\mathbf{y}} 
    \left[ \left\| d_{\mathbf{y}}(s^{+}(e_{\mathbf{x}}(\mathbf{x}))) - \mathbf{y} \right\|^2 \right],
\end{equation}
while $\mathcal{L}_{\mathbf{y}_{\epsilon}\mathbf{x}}$ measures the error of the inverse mapping $s^{-}: \mathcal{Z}_{\mathbf{y}} \to \mathcal{Z}_{\mathbf{x}}$,
\begin{equation}\label{eq:loss_yx}
    \mathcal{L}_{\mathbf{y}_{\epsilon}\mathbf{x}} = \mathbb{E}_{\mathbf{y}_{\epsilon},\mathbf{x}} 
    \left[ \left\| d_{\mathbf{x}}(s^{-}(e_{\mathbf{y}}(\mathbf{y}_{\epsilon}))) 
    - \mathbf{x} \right\|^2 \right],
\end{equation}
The last term $\mathcal{R}$ represents other regularization terms in the loss functions. Here it is used to enforce agreement between the two representations of a given sample in each latent space: the encoded quantities and the quantities obtained by mapping from the other latent space,
\begin{equation}\label{eq:loss_latent}
    \mathcal{R} = \mathbb{E}_{\mathbf{y}_{\epsilon},\mathbf{x}} 
    \left[ \left\| s^{-}(e_{\mathbf{y}}(\mathbf{y}_{\epsilon})) 
    - e_{\mathbf{x}}(\mathbf{x}) \right\|^2  + 
     \left\| s^{+}(e_{\mathbf{x}}(\mathbf{x})) 
    - e_{\mathbf{y}}(\mathbf{y}_{\epsilon}) \right\|^2 \right],
\end{equation}
Without it, nothing constrains the image of the latent mappings to coincide with the image of the encoders: the model could learn one latent geometry for the autoencoding path and a separate one for the mapping path.\\
All loss terms are estimated via mini-batch empirical averages, corresponding to mean squared error minimization, and the model is optimized using the Adam algorithm; further details on the training setup and hyperparameter choices are provided in Sec.~\ref{sec:Dataset_and_training}.

\section{Dataset and training}\label{sec:Dataset_and_training}

\subsection{Synthetic radiances dataset}
\label{sec:title}

\begin{figure}[ht]
    \centering
    \includegraphics[width=12cm]{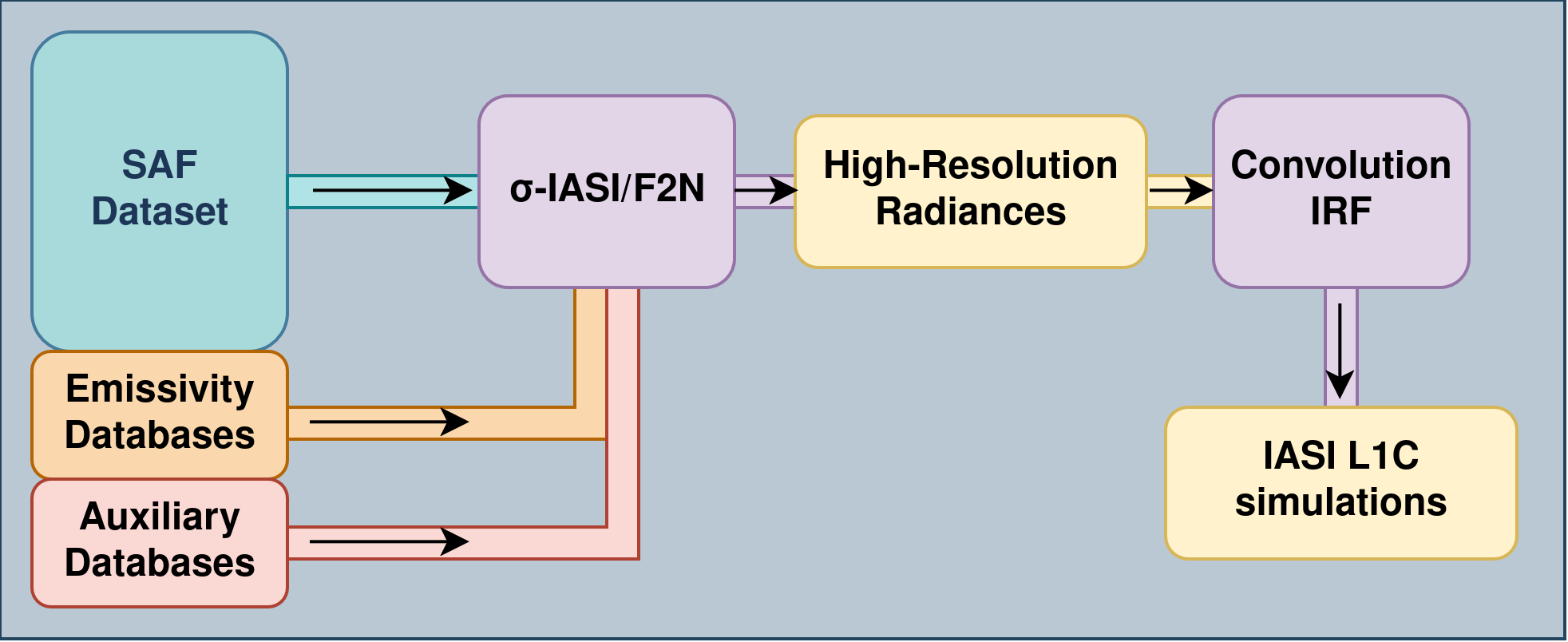}
    \caption{Flow diagram of the complete pipeline for radiance computation.}
    \label{Network}
\end{figure}

The primary dataset used in this study is the diverse profile database from the ECMWF 137-level short-range forecasts \cite{SAF_database}, hereafter referred to as the SAF database. It consists of $25000$ atmospheric profiles derived from global operational short-range forecasts, organized into five subsets designed to provide representative sampling of temperature, specific humidity, ozone mixing ratio, cloud condensates, and precipitation. For each atmospheric scenario, the database provides vertical profiles of temperature, water vapour, and ozone, along with surface properties and ancillary variables such as latitude, longitude, and date. The corresponding top-of-atmosphere clear-sky radiances are computed using $\sigma$-IASI/F2N \cite{MASIELLO2024}, a state-of-the-art fast radiative transfer model capable of simulating all-sky radiances across the spectral range $10$-$2760$ cm$^{-1}$, as illustrated in Fig.~\ref{Network}.
Gas species not included in the SAF database are derived from the IG2 climatological dataset \cite{remedios08}, and surface emissivity is assumed to be Lambertian, derived from the Huang database \cite{huang16}.\\
For each scene, the IASI acquisition is simulated assuming nadir-looking geometry and by convolving the high-resolution radiances with the instrument spectral response function. Finally, each synthetic spectrum is perturbed according to the IASI L1-C measurement noise \cite{serio2020}. Out of $25000$ cases, we perform a random split with $20000$ cases for the training set, $2500$ for the validation set, and $2500$ for the test set.

\subsection{IASI observations dataset}\label{subsec:dataset}

In addition to the synthetic IASI dataset, a set of real IASI Level~1C radiances and the associated Level~2 geophysical retrieval products are also considered~\cite{eumetsat_l1c,eumetsat_l2}. 
In this study, a period of 16 consecutive orbits spanning approximately 24~h of continuous acquisition is considered, as summarized in Table~\ref{tab:Orbits_dataset}.

\begin{table}[H]
    \centering
    \begin{tabular}{@{}lll@{}}
        \toprule
        \multicolumn{3}{c}{\textbf{IASI orbits}} \\
        \cmidrule(r){1-3}
        Start Date & Start Time & Orbit ID \\
        \midrule
        2023-06-30 & 22:29:57 & 24108 \\
        \addlinespace
        End Date & End Time & Orbit ID \\
        \midrule
        2023-07-01 & 23:50:53 & 24123 \\
        \midrule
        \multicolumn{3}{c}{Total Orbits: 16} \\
        \bottomrule
    \end{tabular}
    \caption{Summary of starting and ending IASI orbits used for the analysis. The orbits are consecutive, spanning an acquisition period of approximately 24 hours.}
    \label{tab:Orbits_dataset}
\end{table}

Only nadir-viewing observations were retained, restricting the pixel zenith angle to values not exceeding $8^{\circ}$. This choice minimizes the dependence of the retrieved spectral and geophysical quantities on the viewing geometry, reducing footprint distortion and slant-path atmospheric effects that increase toward the swath edges. 
Clear-sky and cloudy scenes were discriminated using a combination of Level~1 and Level~2 information. 
After applying the geometric, quality, and cloud-screening criteria, the resulting dataset comprises 25817 scans.\\
For each selected observation, the corresponding Level~2 products, vertical profiles of temperature, water vapor, and ozone on pressure levels were retained, along with surface temperature, surface pressure, and spectral emissivity.\\ 

\subsection{Data preprocessing and training setup}

Since only specific quantities of the Earth system are of interest, a channel selection is performed on the IASI spectra. The methodology, first presented for the selection of infrared sounder channels for data assimilation in numerical weather prediction \cite{sgattoni_2025_poster}, combines physical considerations with statistical redundancy analysis. The algorithm first identifies representative super-channels following the concept proposed by McMillin and Goldberg \cite{mcmillin97}, and subsequently removes channels that are highly correlated with these representatives. 
The selection specifically targets the main atmospheric variables, namely atmospheric temperature, water vapor, ozone, and surface temperature, while channels affected by trace gases are discarded from the outset. The radiance is then converted to brightness temperature, to reduce the degree of non-linearity of the problem.\\
In the $\sigma$-IASI/F2N forward model, the number of vertical layers may vary between samples, complicating the construction of a uniformly dimensioned input matrix. To keep the number of vertical layers fixed, we adopt the approach of Sgattoni et al.\cite{Sgattoni_2026}, which can be interpreted as a form of vertical stretching. Starting from the forward model pressure grid, each atmospheric column is re-parameterized onto a normalized vertical coordinate $t \in [0,1]$, where $t = 0$ corresponds to the top of the atmosphere and $t = 1$ to the surface, and is then interpolated onto $60$ layers uniformly spaced over that interval. This procedure is applied to the pressure grid $p$ and to all atmospheric vertical profiles fed into the machine learning architecture, ensuring consistent dimensionality across samples for temperature profile $T$, water vapor profile $w_{\text{vap}}$, and ozone profile $o$. Since the $\sigma$-IASI/F2N forward model uses a fixed
pressure grid, the surface pressure $p_0$ determines the stretching uniquely.\\
We define the observation vector $y \in \mathbb{R}^{m}$, with $m = 944$, as
\begin{equation}
y =
\begin{bmatrix}
y_{ch} & p_0 & p & \text{lon} & \text{lat} & \text{mon}
\end{bmatrix}^\top,
\end{equation}
which includes the simulated noisy IASI radiance brightness temperature for the selected channels, $y_{ch} \in \mathbb{R}^{880}$; the surface pressure, $p_0 \in \mathbb{R}$; the layer pressure grid, $p \in \mathbb{R}^{60}$, obtained via the vertical-stretching procedure described above; the geolocation of the observation, given by longitude and latitude, $\text{lon}, \text{lat} \in \mathbb{R}$; and the month of acquisition, $\text{mon} \in \mathbb{N}$. It is important to note that the SAF database samples atmospheric states that are uncorrelated in space and time, so lat, lon and mon do not identify individual scenes but provide climatological context.\\
Along with the observation vector, we define the atmospheric state vector $x \in \mathbb{R}^{238}$, collecting parameters that describe atmospheric gas constituents and surface properties. 
The state vector is defined as
\begin{equation}
x =
\begin{bmatrix}
TS & T & w_{\text{vap}} & o & \boldsymbol{\varepsilon}
\end{bmatrix}^\top,
\end{equation}
where the main variables described are the Earth's surface temperature, $TS \in \mathbb{R}$, the surface spectral emissivity, $\boldsymbol{\varepsilon} \in \mathbb{R}^{57}$, defined over the spectral interval from $650$ to $2300$ $\text{cm}^{-1}$, on a grid with spacing $10$ $\text{cm}^{-1}$ within the atmospheric spectral windows and $50$ $\text{cm}^{-1}$ outside them. The remaining variables are vertical profiles defined on the same $60$-layer pressure grid $p$ introduced above.\\
The vertical concentrations of water vapor and ozone are mapped through the inverse softplus function
\begin{equation}
\tilde{x} = \ln\!\left(e^{x \cdot c} - 1\right),
\end{equation}
where $x$ is the quantity of interest and $c$ is a scaling constant, set to $c = 1$ for water vapor and $c = 10^6$ for ozone. This transformation is approximately linear for $cx \gg 1$ and approximately logarithmic for $cx \ll 1$, thus compressing the range of magnitudes spanned by the physical quantities and allowing the model to assign sufficient importance, during training, to regions of very low concentration, such as stratospheric water vapor and tropospheric ozone.\\
Both the observation and state vectors are then normalized using min-max normalization, rescaling each variable to the range $[0,1]$.\\
The model was trained using the Adam optimizer \cite{kingma2017}. 
The latent dimensions were set to $\dim(\mathcal{Z}_\mathbf{x}) = 100$ and $\dim(\mathcal{Z}_\mathbf{y}) = 100$, corresponding to compression factors of approximately 2.4 and 9.4 with respect to the state and observation spaces, respectively. These values, together with the encoder and decoder widths and the loss weights $\omega$, were selected on the validation set.
All loss terms are estimated via mini-batch empirical averages, corresponding to mean squared error (MSE) minimization. A batch size of $64$ was used, and dropout regularization with a rate of $0.05$ was applied to mitigate overfitting. During training, a reduce-on-plateau scheduling policy was adopted: the initial learning rate of $10^{-4}$ was reduced by a factor of $0.1$ whenever the validation loss did not improve for $50$ consecutive epochs, allowing the optimizer to take progressively finer steps as it approached a minimum.

\section{Results}

\subsection{Application to synthetic acquisitions}\label{sec:synthetic}

\begin{figure} [ht]
\begin{center}
\begin{tabular}{c} 
\includegraphics[height=5.5cm]{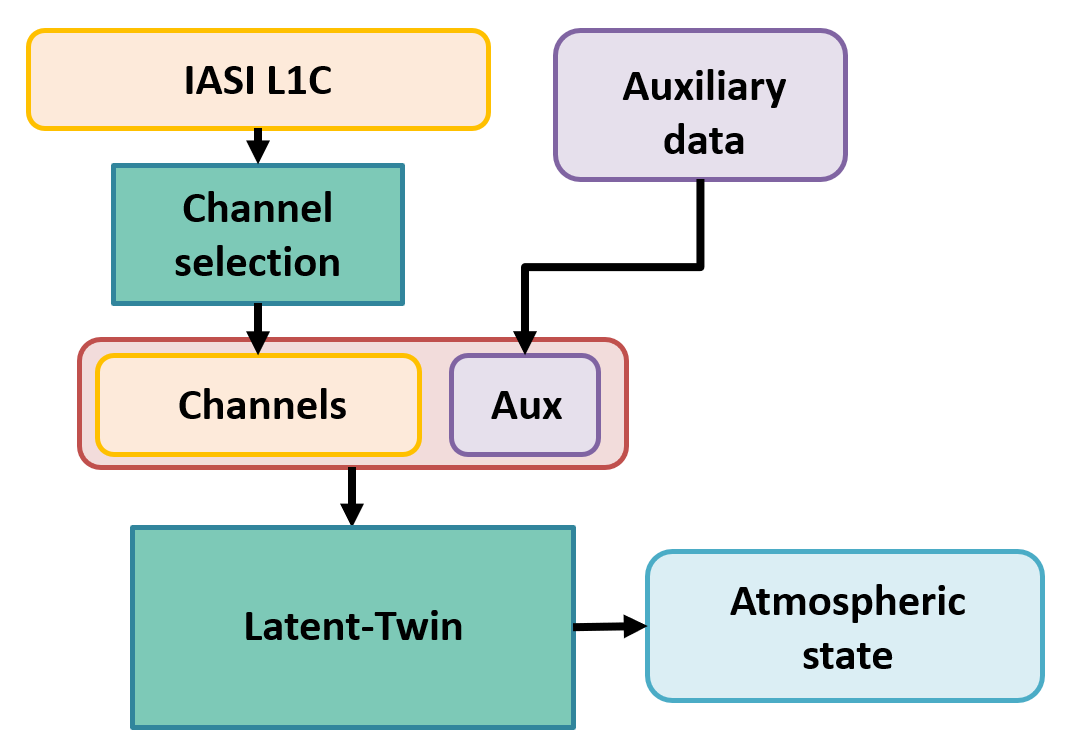}
\end{tabular}
\end{center}
\caption[LTarchitecture] 
   { \label{fig:pipeline_input} Schematic of the input-generation process for the latent twins architecture. IASI-L1C radiance channels are selected and converted to brightness temperature before being inserted into the vector $y$ (red box), together with auxiliary data (geolocation, time, and pressure).
}
\end{figure}

To perform an initial consistency test and evaluate possible model biases, we applied the inversion routine to the synthetic radiance test set. Fig.~\ref{fig:pipeline_input} shows the pipeline of the input-generation process for the latent twins architecture.

\begin{figure}[ht]
	\centering
	\includegraphics[width=13cm]{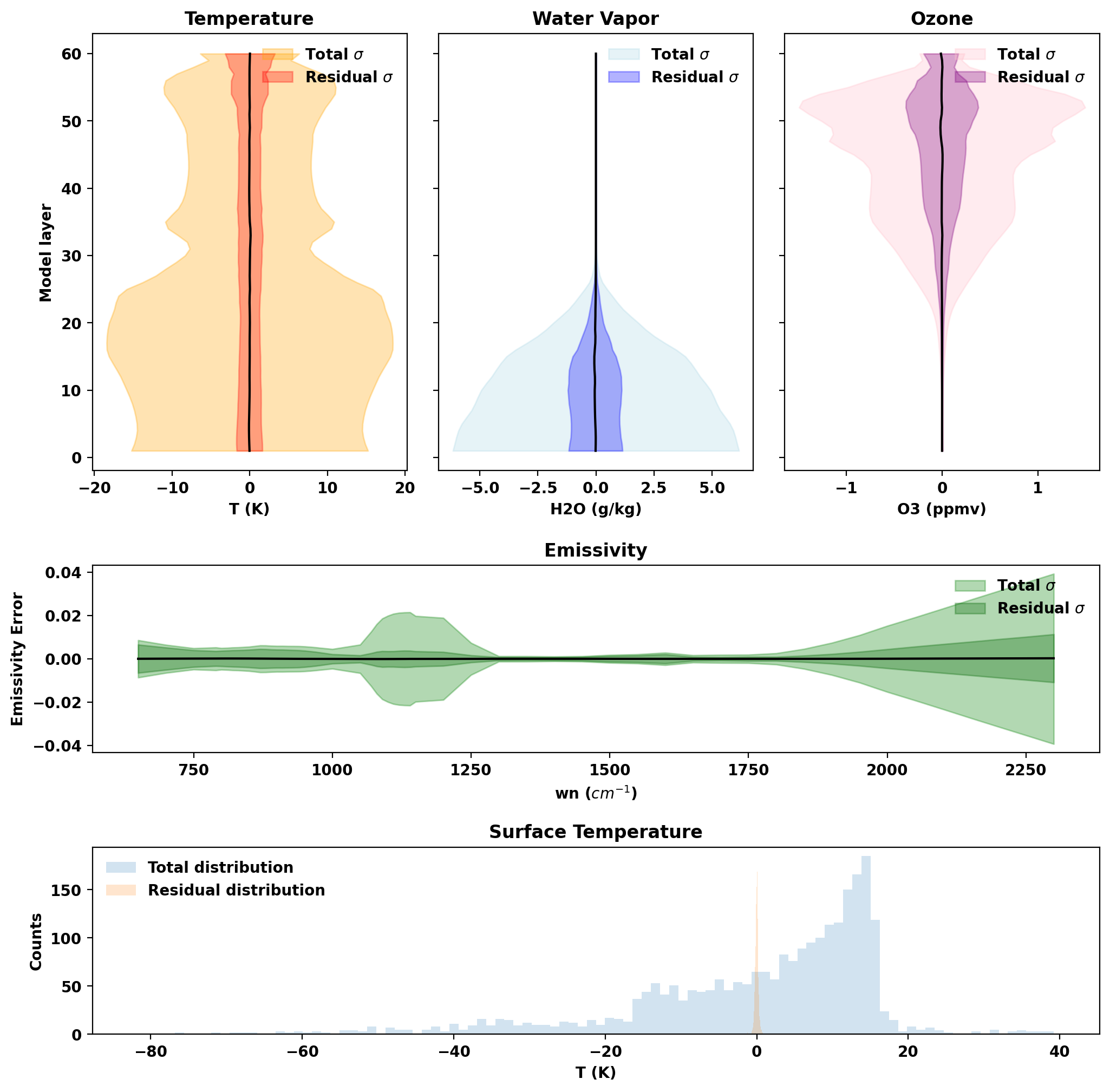}
	\caption{Vertical profiles and spectral behaviour of the retrieval residuals over the test set. The top row shows, for temperature, water vapor, and ozone, the Mean Bias Error (black line) as a function of model layer, together with the standard deviation ($\sigma$) of the residuals $x_{res}$ and of the test-set deviation $\Delta x$ (shaded areas). Similarly, the middle panel compares the residuals for the emissivity spectrum. The bottom panel shows the distribution of surface temperature residuals (orange) compared to the total distribution of the test-set deviations (blue).}\label{fig:residuals_test}
\end{figure}

The residual of the retrieved quantities with respect to the true values is computed as:

\begin{equation}
    x_{res} = x - x_{true},
\end{equation}
while the standard deviation of the test set is computed from the deviation of the true quantity from its mean:

\begin{equation}
    \Delta x = x_{true} - \left< x_{true} \right>,
\end{equation}
The results are shown in Fig.~\ref{fig:residuals_test}. For the temperature, water vapor, ozone, and emissivity profiles, the black line indicates the Mean Bias Error (MBE), while the shaded areas represent the standard deviation of the residuals $x_{res}$ and of the test-set deviation $\Delta x$. The distribution of the surface temperature residuals is shown separately.

\begin{figure}[ht]
	\centering
	\includegraphics[width=10cm]{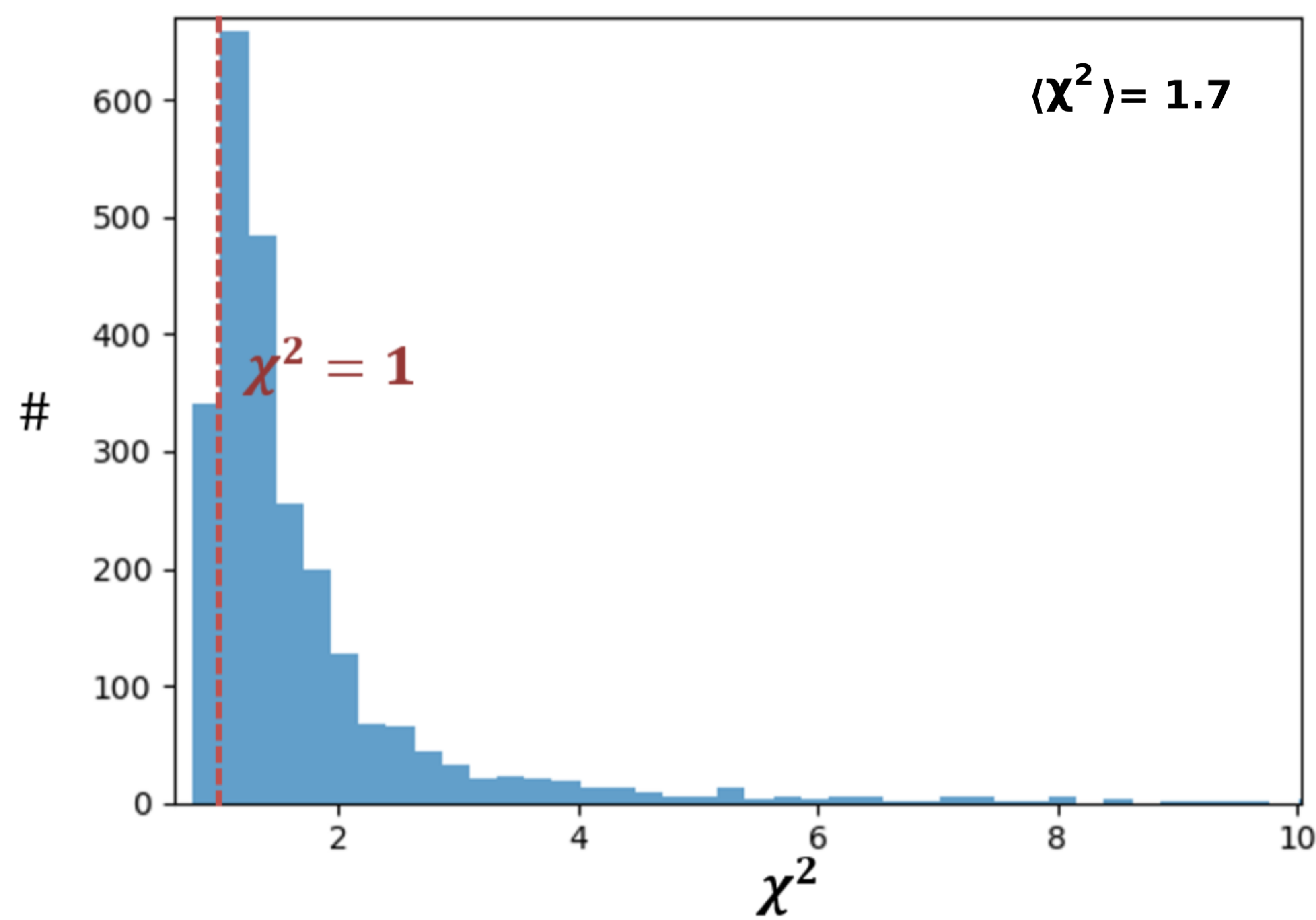}
	\caption{Reduced $\chi^2$ distribution computed over the test set.}\label{fig:chi2_test}
\end{figure}

The results indicate an unbiased retrieval, with standard deviation errors of approximately $1.6$~K for tropospheric temperature, $1.1$~g/kg for tropospheric water vapor, $0.3$~ppmv for stratospheric ozone, $1\%$ for emissivity, and $0.22$~K for surface temperature.
These quantities correspond to the residual statistics over the test set, and describe a climatological error over the distribution sampled by the SAF database. It is possible, as a first approximation, to adopt these quantities as a first-order error characterization of the retrieval however, it is important to note that these quantities are not scene-dependent uncertainty analogous to the retrieval error covariance provided by optimal estimation schemes.\\
To assess the self-consistency of the retrieved quantities with respect to the forward problem, we computed the reduced $\chi^2$ distribution for the residuals between the original simulated radiances and the radiances recomputed by the forward model using the retrieved profiles. This distribution is shown in Fig.~\ref{fig:chi2_test}.

It is important to note that, for a general deep learning architecture, the expected reduced $\chi^2$ distribution is not guaranteed to follow the theoretical distribution with an expectation value of $1.0$. This is because, when the architecture is purely data-driven, the algorithm learns to return the profiles according to the rule imposed by the loss function during training. These profiles are closely related to, but not identical to, the ones that truly produced the observed radiances. Such differences can introduce inconsistencies between the observed and the derived radiances. The latent twins architecture, by contrast, is able to approximate both the forward and inverse processes through a shared latent mapping, thereby informing the inverse problem with the forward problem. This helps produce more radiance-consistent results. The results highlight a reasonable radiance consistency with average $\chi^2 \sim 1.7$.

\subsection{Application to IASI measurements}

\begin{figure} [ht]
\begin{center}
\begin{tabular}{c} 
\includegraphics[height=13cm]{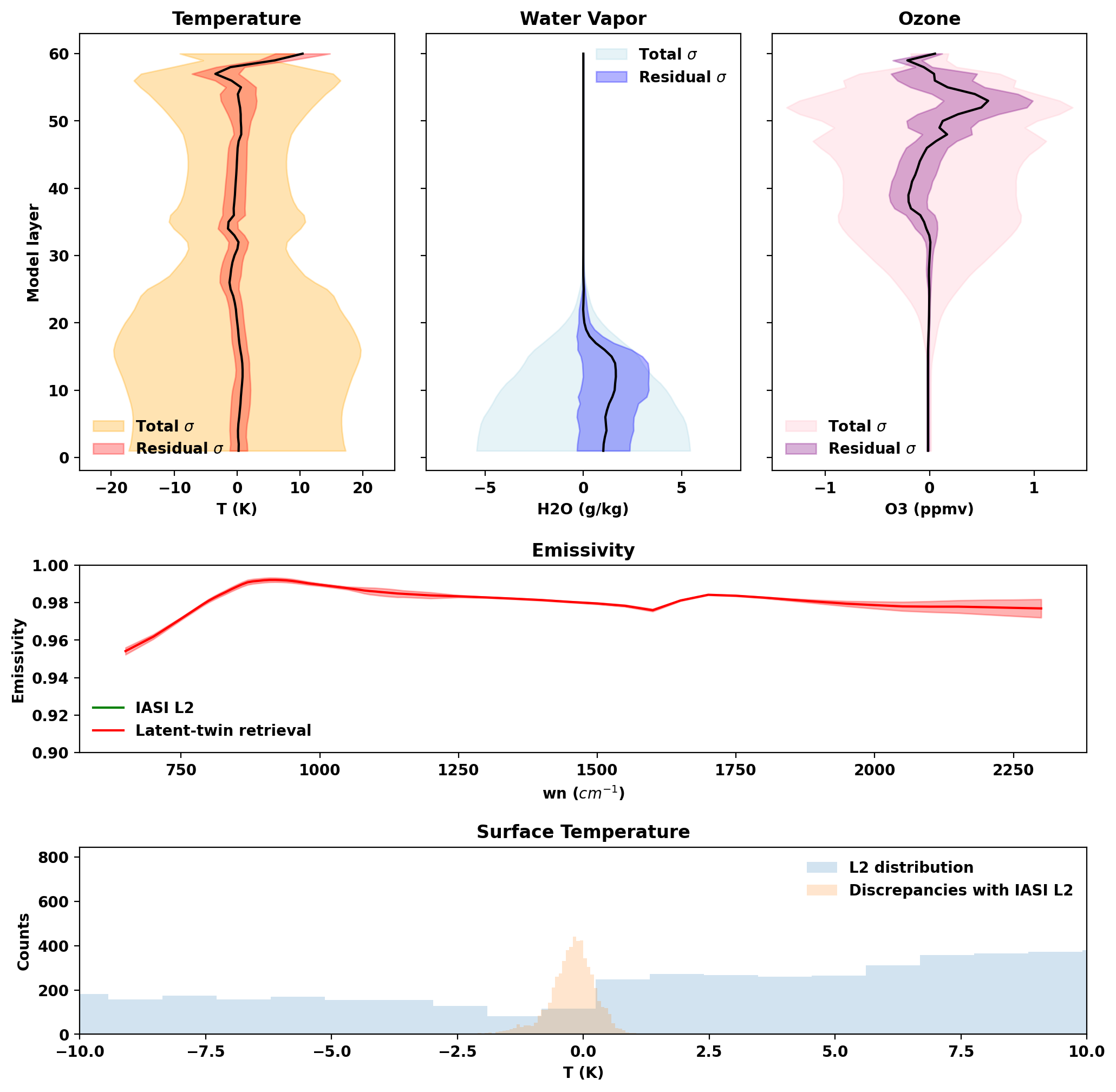}
\end{tabular}
\end{center}
\caption[LTarchitecture] 
   {Comparison between retrieved quantities and IASI-L2 products over sea. The top row shows, for temperature, water vapor, and ozone, the Mean Bias Error (black line) as a function of model layer, together with the standard deviation of the residuals and of the L2 profile (shaded areas). The middle panel shows the mean latent twins retrieved emissivity over sea, as IASI-L2 emissivity is not available for this surface type. The bottom panel shows the distribution of surface temperature residuals (orange) compared to the total distribution of the IASI-L2 surface temperature deviations (blue).\label{fig:L2_residuals_sea} }
\end{figure} 

\begin{figure} [ht]
\begin{center}
\begin{tabular}{c} 
\includegraphics[height=13cm]{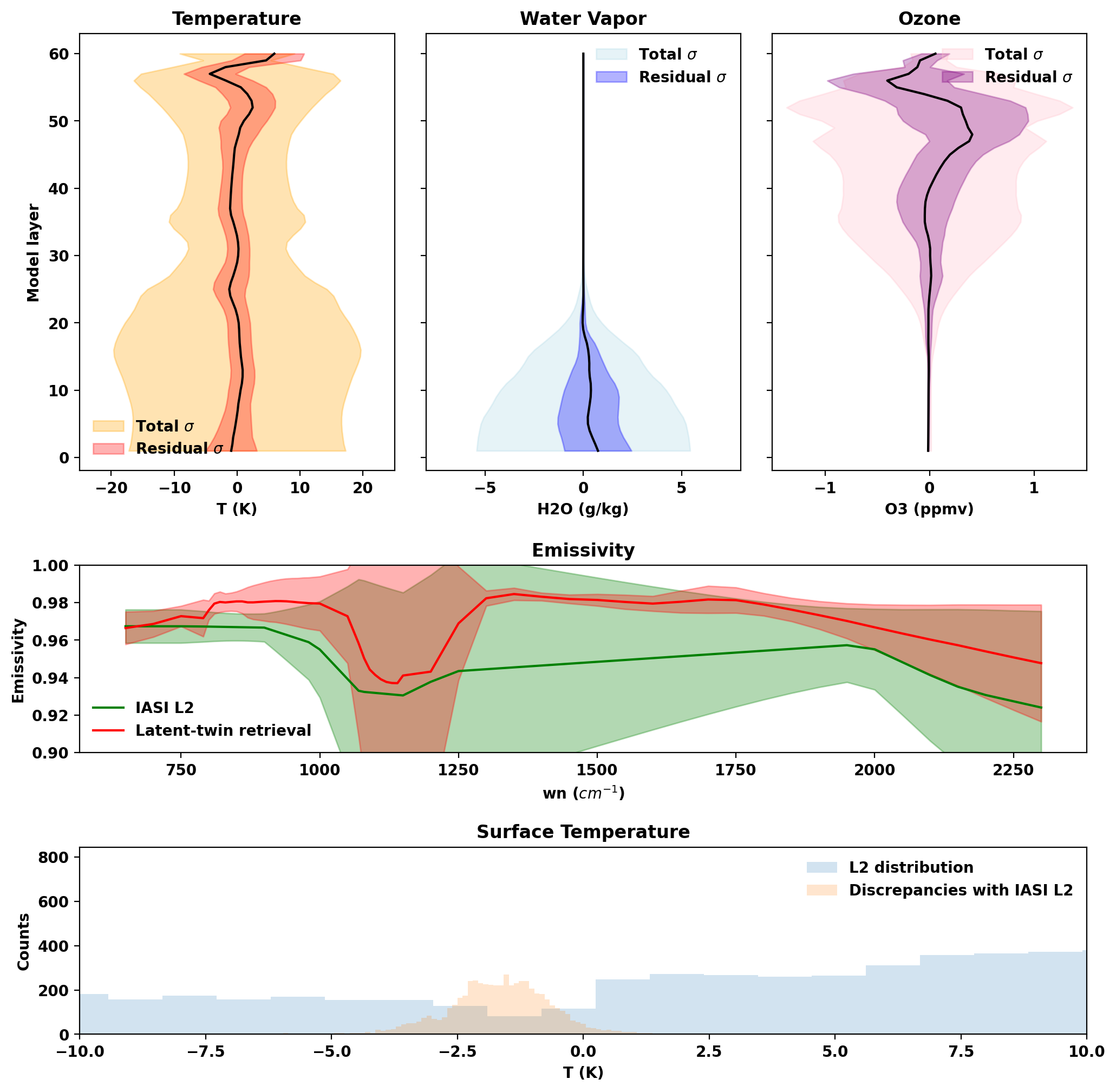}
\end{tabular}
\end{center}
\caption[LTarchitecture] 
   {Same as Fig.~\ref{fig:L2_residuals_sea} but for acquisitions over land. For these scenarios, IASI-L2 emissivity is available, and its average value and standard deviation are shown as a green line and shaded area, respectively, in the second panel.\label{fig:L2_residuals_land} }
\end{figure} 

After validating the method on synthetic data, we assessed its performance on real measurements acquired by the IASI instrument \cite{SIMEONI1997} onboard the Metop satellites. We present results obtained for the selected IASI Level 1C observations, with the corresponding Level 2 products serving as a reference for evaluating the retrieval accuracy.
As in Sec.~\ref{sec:synthetic}, the IASI-L1C radiances are converted to brightness temperature and the channel selection is applied. The input vector is then completed with ancillary information on acquisition geolocation, month, pressure profile and surface pressure.
The retrieved profiles of temperature, water vapor, and ozone are compared with the corresponding IASI-L2 products, projected onto the latent-twin vertical pressure grid $p$.\\
The analysis is conducted by separating the dataset into two groups: one for acquisitions over sea and one for acquisitions over land.
Results for the first set are summarized in Fig.~\ref{fig:L2_residuals_sea}. Temperature profiles agree well between the two products, with no evident systematic differences and a standard deviation of $1.5$~K. Larger discrepancies are found for tropospheric water vapor (about $1.0$~g/kg) and stratospheric ozone ($0.4$~ppmv). Surface properties are also consistent: the retrieved emissivity spectrum matches that of water, and surface temperature agrees well with IASI-L2.\\
Several factors may contribute to these differences. First, the two retrievals rely on different prior information. Although the latent twins does not employ an explicit prior in the Bayesian sense, the minimization of the mean squared error in Eq.~(\ref{eq:loss_yx}) drives the inverse surrogate towards the conditional mean $\mathbb{E}[\mathbf{x} \mid \mathbf{y}]$ evaluated over the training distribution, while the reduced dimensionality of $\mathcal{Z}_\mathbf{x}$ constrains the output to the manifold learned from the SAF database. The training set therefore acts as an implicit prior, whose influence becomes dominant where the measurement information content is low. Since IASI-L2 relies on a different background, the discrepancy between the two priors propagates directly into the comparison. Second, the two retrievals adopt different spectroscopic databases, which differ in their spectroscopic line parameters and continuum formulations. Third, the two products are characterized by different vertical resolutions, so that the comparison is affected by the differing degrees of vertical smoothing.
Finally, it should be noted that IASI-L2 is itself a retrieval with its own error, and the differences reported here are the combined discrepancy between the two products rather than the error of either one.\\
Results for acquisitions over land are presented in Fig.~\ref{fig:L2_residuals_land}. Here the differences in the vertical profiles are less pronounced, with compatible results for temperature, water vapor, and ozone. Surface properties, however, show some discrepancies: emissivity is higher than in the IASI-L2 retrieval, while surface temperature is lower. This compensating behaviour is consistent with the well-known difficulty retrieval systems encounter in disentangling surface temperature from surface emissivity. The emissivity training data may also contribute: the latent-twin model is trained exclusively on Lambertian spectra from the Huang database, so the retrieved emissivity is expected to stay close to the range of variability represented there, which may differ from the one adopted in IASI-L2 processing.

\section{Conclusions}

The framework provides a possible pathway for retrieving atmospheric properties under clear-sky conditions, allowing the simultaneous retrieval of temperature, water vapor, ozone, surface temperature and surface emissivity. When applied to synthetic data, the retrieval is self-consistent, with an average reduced $\chi^2 \simeq 1.7$ indicating a reasonable radiance consistency. On the other hand, the comparison with IASI-L2 products yields a standard deviation of $1.5$~K for temperature and larger differences for water vapor, ozone and land surface properties. A key strength of the approach is its computational efficiency: once trained, the model performs approximately $4\times10^{5}$ retrievals per minute on a standard laptop, against the $10^{2}$-$10^{4}$~s per scene typical of optimal estimation schemes.\\
As discussed in Sec.~\ref{sec:synthetic}, the error characterization adopted here relies on test-set residual statistics and is therefore climatological in nature. It is likely to be optimistic for atmospheric states that are poorly represented in the training database. Proper uncertainty quantification, via variational autoencoders or ensemble estimates, and validation against independent measurements such as radiosonde profiles are left for future work. Further developments include a more extensive analysis on different seasons, aerosol scattering layers and the retrieval of trace gases; the extension to cloudy conditions is addressed in Sgattoni et al.\cite{Sgattoni_2026}.

\section{Data availability}
The training and test datasets used in this study, together with the architecture, are available on request.

\newpage
\bibliography{report} 
\bibliographystyle{spiebib} 

\end{document}